\documentclass[superscriptaddress,twocolumn]{revtex4}

\usepackage{amssymb}
\usepackage{amsfonts}
\usepackage{amsmath}
\usepackage{graphicx}
\usepackage{dcolumn}
\usepackage{bm}
\usepackage[latin1]{inputenc}
\begin{document}
\title{Matching tetrads in $f(T)$ gravity}

\author{Franco Fiorini}
\email{francof@cab.cnea.gov.ar} \affiliation{Departamento de Ingeniería en Telecomunicaciones, Consejo Nacional de Investigaciones Científicas y Técnicas (CONICET) and Instituto Balseiro (UNCUYO), Centro Atómico Bariloche, Av. Ezequiel Bustillo 9500, CP8400, S. C. de Bariloche, Río Negro, Argentina.}

\author{Martín Onetto}
\email{martin.onetto@ib.edu.ar}
\affiliation{Sección Física Forense, Consejo Nacional de Investigaciones Científicas y Técnicas (CONICET) and Instituto Balseiro (UNCUYO), Centro Atómico Bariloche, Av. Ezequiel Bustillo 9500, CP8400, S. C. de Bariloche, Río Negro, Argentina.}

\begin{abstract}
The procedure underlying the matching of 1-form (tetrad) fields in theories possessing absolute parallelism --$f(T)$ gravity being within this category-- is addressed and exemplified. We show that the \emph{remnant} symmetries of the intervening spaces play a central role in the process, because the knowledge of the remnant group of local Lorentz transformations enables one to perform rotations and/or boosts in order to $\mathcal{C}^{1}$-match the corresponding tetrads on the junction surface. This automatically ensures the continuity of the Weitzenb\"{o}ck scalar there, even though this proves to be just a necessary condition in order to obtain a global parallelization of the spacetime.
\end{abstract}
\maketitle

\section{Introduction}

Junction conditions are key tools in regard to the comprehension on how different solutions of a given physical theory interact. They provide crucial information about the physical constraints arising because of the interplay between the intervening solutions. In fact, the way the pieces combine into a more global solution is as important as the individual pieces themselves. For instance, vacuum solutions describing black holes in any gravitational theory would be of little interest if we did not know that black holes are the result of gravitational collapse, i.e., of the unstoppable contraction of matter in a certain region of spacetime. In this case, junction conditions will not only be instrumental in providing a black hole formation mechanism, but also in relating the physical properties of the collapsing matter to the angular momentum, electric charge, and mass of the remaining black hole.

In General Relativity (GR) the definite answer to the junction problem was pronounced by Israel \cite{Israel}, even though Darmois and, later on, Lichnerowicz, were very aware of the issue and made sharp contributions on the matter well before \cite{lake}. Actually, solutions of GR involving matching techniques were known since the late thirties and early forties, especially concerning the junction between Friedmann-Robertson-Walker cosmological models and the spherically symmetric vacuum corresponding to the Schwarzschild solution \cite{Opp},\cite{Strauss}. Much more recently, the junction conditions in extended theories of gravity within the metric approach --known as $f(R)$ gravity-- were worked out \cite{natalie},\cite{Senovilla}. Additionally, junction conditions in Palatini-$f(R)$ gravity were studied as well \cite{Gonzalo}.

Here it is our intention to study junction problems at the level of the tetrad field, which is the dynamical carrier of the gravitational degrees of freedom in the so called $f(T)$ gravity \cite{FF}-\cite{Ferraro:2011}. Due to the additional complications coming from the underlying parallelization process involved in these theories, it is by no means surprising that very few authors dealt with this subject in the past, see \cite{de la cruz},\cite{deben}. In certain sense we offer in this article a complementary point of view regarding the ideas developed in \cite{de la cruz}; actually, by bringing into scene the remnant group of local Lorentz symmetries present in $f(T)$ gravity, we explain by means of a relatively simple example, how to get a global parallelization which incorporates at once the partial ones associated to the spaces to be glued. This procedure is particularly useful when no matter shells or layers are present on the junction surface, so it can be considered the $f(T)$ analogue of the Darmois-Lichnerovicz procedure. Hopefully it will serve also in other theories relying on absolute parallelism (see, e.g. \cite{Vatu}), even though a characterization of the remnant symmetries in question constitutes a crucial ingredient of the method.

After a brief exposition of $f(T)$ gravity and its remnant group of transformations in section \ref{Chpt:BlackHol}, we heuristically illustrate the nature of the matching problem in sec. \ref{sec2}. Section \ref{the matching} exemplifies the technique by obtaining a global parallelization for the Schwarzschild interior and exterior spaces corresponding to a simple model of a spherical star immersed in a spherically symmetric vacuum. Finally, we further comment on our results in \ref{remarks}.

\section{Preliminary material}
\label{Chpt:BlackHol}

In the context of $f(T)$ gravity the dynamical field is the tetrad $e^a(x^\mu)$, and the metric $\textbf{g}(x^\mu)$ is just a subsidiary field obtained by means of $\textbf{g}(x^\mu)=e^a(x^\mu)e^b(x^\mu)\eta_{ab}$, where $\eta_{ab}=diag(1,-1,-1,-1)$. This means that in the local coordinates $x^\mu$, the components of the metric are just $g_{\mu\nu}=e^a_{\,\,\mu}e^b_{\,\,\nu}\eta_{ab}$. The tetrad is responsible for the torsion $T^{a}=de^a$, which in local components acquires the form $T^{a}{}_{\mu\nu}=\partial_{\mu} e^{a}_{\,\,\nu}-\partial_{\nu} e^{a}_{\,\,\mu}$. According to this framework, the tensor $T^{a}$ is the building block of the gravitational lagrangian; in fact, we can construct the (diffeomorphism) invariant
\begin{equation}\label{escalar}
T = S^{a}{}_{\mu\nu} T_{a}{}^{\mu\nu},
\end{equation}
where
\begin{align*}
  S^{a}{}_{\mu\nu} = \frac{1}{4}
  (T^{a}{}_{\mu\nu} - T_{\mu \nu }{}^{a} + T_{\nu \mu }{}^{a}) +
  \frac{1}{2} (\delta_{\mu}^{a} T_{\sigma\nu}{}^{\sigma} -
  \delta_{\nu}^{a} T_{\sigma\mu}{}^{\sigma})\,,
\end{align*}
and $T_{\sigma }{}^{\sigma\nu}=g^{\mu\rho}T_{\mu\rho }{}^{\nu}=e^{\mu}_{\,\,a}e^{\rho}_{\,\,b}\eta^{ab}T_{\mu\rho }{}^{\nu}$. The \emph{Weitzenb\"{o}ck} invariant $T$ is a scalar only under general coordinate transformations. However, under a local Lorentz transformation of the tetrad, $e^{a}(x^\mu)\rightarrow \tilde{e}^{a}(x^\mu)=\Lambda^{a}_{\,\,b}(x^\mu)e^b(x^\mu)$, the object $T$ transforms as $T\rightarrow \tilde{T}=T+DT$, with

\begin{equation}\label{transft}
DT~=~e^{-1}d(\epsilon_{abcd} \,e^{a}\wedge e^{b}\wedge \eta^{de} \Lambda^{c}_{\,\,f}d\Lambda^{f}_{\,\,e}),
\end{equation}
and $e=\det e^a_{\,\,\mu}=\sqrt{|\det g_{\mu\nu}|}$. This transformation law has a central importance in this work; we see that, in general, only global Lorentz transformations leave $T$ invariant. Nonetheless, we shall briefly comment on the local symmetries which actually lie behind the transformation law encoded in (\ref{transft}), leaving the reader to follow the full exposition given in \cite{grupo} and \cite{NosUn}.

The \emph{remnant group} of local Lorentz transformations \cite{grupo}, is defined by demanding that $T$ becomes a genuine Lorentz scalar, i.e., by asking the on-shell equation $DT=0$. The remnant group can be characterized by expanding the Lorentz transformations around the identity element, i.e.,

\begin{equation}\label{matlor}
\Lambda^{a}_{\,\,b}(x^\mu)=\delta^a_b +\frac{1}{2} \sigma^{cd}(x^\mu)(M_{cd})^{a}_{\,\,b}+\mathcal{O}(\sigma^2),
\end{equation}
where $\sigma^{cd}(x^\mu)=-\sigma^{dc}(x^\mu)$ are the infinitesimal parameters and

\begin{equation}\label{algebra}
(M_{cd})^{a}_{\,\,b}=\delta^a_c\,\eta_{db}-\delta^a_d\,\eta_{cb}.
\end{equation}
It is direct to show that the requirement $DT=0$ coming from (\ref{transft}), at the same order implies

\begin{equation}\label{condinfi}
\epsilon_{abcd} \, d(e^{a}\wedge e^{b})\wedge d\sigma^{cd}+\mathcal{O}(\sigma^2)=0.
\end{equation}
This result seems to suggest the convenience of classifying a given spacetime $e^a$ in terms of the closed 2-forms $e^a\wedge e^b$ it involves, if any; in this way, a given tetrad will be called a $n$-closed-area frame ($n$-CAF), if it satisfies $d(e^{a}\wedge e^{b})=0$ for any of the six different pairs $(a,b)$ ($0\leq n\leq6$). For instance, let us inquire into the infinitesimal, local symmetries allowed by the 6-CAF \emph{Euclidean}, tetrad $e^{a}=\delta^{a}_{b}dx^{b}$, which leads to Minkowski space with metric $\eta_{ab}$. Eq. (\ref{condinfi}) just says that the infinitesimal parameters $\sigma^{cd}$ are totally free. This means that any non-local Lorentz invariant theory constructed upon the Weitzenb\"{o}ck invariant $T$, such as $f(T)$ gravity to be described in a moment, is unable to select a preferred parallelization at a local level. In other words, the absence of gravity embodied in Minkowski space is described in $f(T)$ gravity as a local lack of preferred frames; any of them connected by local, infinitesimal Lorentz transformations is equally good for describing the local absence of gravity. However, things are very different for finite Lorentz transformations. In this case the full expression (\ref{transft}) must be considered, and one can prove that six one-dimensional groups (three pure rotations and three pure boosts) act as remnant symmetry groups of Minkowski space. Moreover three Abelian, two-dimensional groups $G_{i}$ appear as symmetries as well; these are generated by $G_{i}=\{K_{i},J_{i}\}$, $(i=x^1,x^2,x^3)$, where $K_{i}$ and $J_{i}$ correspond to boosts and rotations along and about the $i$-axes,respectively. For details, we suggest to consult \cite{grupo}.


The equations of motion of $f(T)$ gravity are

\begin{eqnarray}
  \bigl[e^{-1}\partial_\mu(e\ S_a{}^{\mu\nu})+e_a^\lambda T^\rho{}_{\mu\lambda} S_\rho{}^{\mu\nu}\bigr]
  f^{\prime }+&\nonumber\\
  +\,S_a{}^{\mu\nu} \partial_\mu T f^{\prime \prime } - \frac{1}{4}
  e_a^\nu f =& -4\pi G e_a^\lambda \mathcal{T}_\lambda{}^{\nu},
  \label{ecuaciones}
\end{eqnarray}
where an arbitrary (at least twice differentiable) function $f$ of the Weitzenb\"{o}ck invariant $T$ appears. The equations are derived from the action
\begin{align}
  I=\frac{1}{16 \pi G }\int f(T)\,  e\, d^4x+I_{\rm matter}\,,
  \label{actionfT3D0}
\end{align}%
which reduces to Einstein's General Relativity (in its teleparallel equivalent form), when $f(T)=T$. In (\ref{ecuaciones}), $\mathcal{T}_\lambda{}^{\nu}$ is the energy momentum tensor coming from the matter action $I_{\rm matter}$, and primes denote differentiation with respect to $T$. Note that the equations of motion are of second order in derivatives of the tetrad; this is a consequence of the fact that $T$ is constructed upon first derivatives of the tetrad. Then, a proper set of initial data is $S=\{e^a,\partial_{\mu}e^a\}$, and different solutions should be matched at junctions surfaces $J$ in such a way that the tetrad and its first derivatives are well defined on them. The differentiability conditions on $J$ automatically assure the existence of the torsion and the Weitzenb\"{o}ck invariant on $J$. This is slightly different from GR, where the scalar curvature $R$ does not need to be continuous on $J$ because of the fact that it contains second derivatives of the metric.

\section{Heuristics}\label{sec2}

In this section we shall illustrate the nature of the problem by a simple example which, nonetheless, captures the essence of the matching procedure in the presence of absolute parallelism. Let us consider Minkowski spacetime $\mathcal{M}$ described in standard pseudo-Euclidean coordinates $(t,\textbf{x})$ as the union of the submanifolds $\mathcal{M}= \bigcup_{i}\mathcal{M}_{i}$, where

\begin{eqnarray} \label{division}
\mathcal{M}_{1}&=&\{(t,\textbf{x})\in \mathcal{M}\,|\, t\geq1\},\notag\\
\mathcal{M}_{2}&=&\{(t,\textbf{x})\in \mathcal{M}\,|\, -1<t<1\},\notag\\
\mathcal{M}_{3}&=&\{(t,\textbf{x})\in \mathcal{M}\,|\, t\leq-1\}.
\end{eqnarray}
Here the Cauchy hypersurfaces $t=\pm 1$ will play the role of junction surfaces $J_{\pm}$. The spacetime metric is simply $\eta_{ab}$ everywhere, regardless of the region considered. However, the $f(T)$ equations of motion are solved by infinitely many 1-form fields representing the infinite ways of parallelizing $\mathcal{M}$. Let us suppose that in solving the equations, we find the fields

\begin{equation}\label{tetradapartidabis1}
e^{a}=\delta^{a}_{b}dx^{b},\,\,\,\, \mbox{in $\mathcal{M}_{1}$ and $\mathcal{M}_{3}$}.
\end{equation}
As mentioned in the previous section, the remnant group of $\mathcal{M}$ includes 6 one-dimensional groups of boosts and rotations. Suppose that in solving the equations in $\mathcal{M}_{2}$, the following $x$-boosted tetrad is found:

\begin{equation}\label{tetradapartidabis2}
                             \begin{array}{l}
                              e^{0}=\cosh[\theta(t,\textbf{x})]dt+\sinh[\theta(t,\textbf{x})]dx \\
                              e^{1}=\sinh[\theta(t,\textbf{x})]dt+\cosh[\theta(t,\textbf{x})]dx \\
                              e^{2}=dy \\
                              e^{3}=dz
                              \end{array},\,\,\,\,
                              \mbox{in $\mathcal{M}_{2}.$}
                               \end{equation}

Despite the manifestly artificial character of the splitting given in (\ref{division}), the frames (\ref{tetradapartidabis1}) and (\ref{tetradapartidabis2}) are genuine solutions (in the corresponding regions) of the vacuum $f(T)$ motion equations leading to the metric $\eta_{ab}$; they are connected by a time-dependent boost in the $t-x$ plane, which is a transformation belonging to the remnant group of $\mathcal{M}$. As a matter of fact, the Weitzenb\"{o}ck invariant coming from (\ref{tetradapartidabis1}) and (\ref{tetradapartidabis2}) is identically null, even though the frame in $\mathcal{M}_{2}$ generates several non null components of the torsion tensor, namely

\begin{eqnarray} \label{comptor}
T^{txt}&=&\partial_{t}\theta,\notag\\
T^{txi}&=&\partial_{i}\theta,\,\,\,i:y,z\notag\\
T^{xit}&=&\partial_{i}\theta,\,\,\,i:x,y,z.
\end{eqnarray}
However, the fields (\ref{tetradapartidabis1}) and (\ref{tetradapartidabis2}) do not conform a parallelization of the whole space $\mathcal{M}$ in general, because of the discontinuity produced in the junction surfaces $J_{\pm}$. Plainly the tetrads are not $C^{1}$-matched there.

In this trivial example, things can be solved easily. Using the freedom to choose the tetrad provided by the remnant group, we can sort things out on the junction surface. For instance, taking $\theta(t,\textbf{x})=0$ on $\mathcal{M}_{2}$, a global parallelization of $\mathcal{M}$ is obtained at once, leading to the canonical, \emph{Euclidean} frame $e^{a}=\delta^{a}_{b}dx^{b}$ everywhere. Of course, this is the simplest and more diaphanous solution concerning the parallelization of $\mathcal{M}$ from the outset, but infinitely many other, perhaps more interesting choices can be made; among them, we have $\theta(t,\textbf{x})=(1+cos(\pi\,t))/2$, which offers a $\mathcal{C}^1$ matching of the tetrads on the surfaces $J_{\pm}$, ensuring thus the continuity of the torsion tensor on them.

The simple setting involved in this example teaches us three valuable lessons, which are as follows:
\begin{enumerate}
  \item What is obvious at the level of the metric (in this case $\eta_{ab}$), is not in the underlying world of the tetrad $e^a$.
  \item The continuity of the Weitzenb\"{o}ck invariant $T$ on the junction surface is just a \emph{necessary} condition in order to have a well-defined parallelization on $\mathcal{M}$. A sufficient condition is given by the continuity of the first derivatives of $e^a$ on the junction surface or, equivalently, of the components of the torsion $T^a$ on it. These facts are in agreement with the analysis performed in \cite{de la cruz}.

  \item The importance of a thorough characterization of the remnant group of the space-time under consideration; this will enable one to select local Lorentz rotations and boosts near the junction surface in order to $\mathcal{C}^1$-match the corresponding tetrads.
\end{enumerate}
In general junction problems, the role of the $\mathcal{M}_{i}$ above could be played by spaces of very different geometrical and topological structure, and it is easy to foresee that things will rapidly go out of control, for in addition to the usual complications arising as a consequence of gluing the spaces at the level of the metric field, the remnant symmetries to be used for matching the tetrads will manifest in a very dissimilar tetrad structure on either side of the junction surface. However, the continuity of $T$ on the junction surface will serve as a guiding principle; once assured, a further knowledge of the remnant group of the intervening spaces will hopefully enable us to perform the $\mathcal{C}^1$-matching at the end. We illustrate the procedure in the next section by using a simple, though physically relevant, model of a star.

\section{Matching tetrads}\label{the matching}

Let us consider the very well-known, textbook, Schwarzschild interior and exterior metrics \cite{Wald}, \cite{clarke},

\begin{equation}\label{schain}
ds^{2}_{-} = \frac{\left[3h(r_{0})-h(r)\right]^{2}}{4} dt^{2} - \frac{dr^{2}}{h(r)^{2}} - r^{2} d\Omega^{2},\,\,\,0\leq r\leq r_{0},
\end{equation}
where $h(r) = \left(1-r^{2}/R^{2}\right)^{1/2}$, $R^{2} = 3/(8\pi \varrho_{0})$, and
\begin{equation}\label{schaout}
ds^{2}_{+} = (1-2M/r) dt^{2} - \frac{dr^{2}}{1-2M/r} - r^{2} d\Omega^{2},\,\,\,\,r\geq r_{0}.
\end{equation}
Metric (\ref{schain}) represents a simple model of a static spherically symmetric star of radius $r_{0}$, consisting of a perfect fluid with constant energy density $\varrho_{0}$ and pressure
\begin{equation}\label{presion}
p(r) = \varrho_{0}\left(\frac{h(r)-h(r_{0})}{3h(r_{0})-h(r)}\right).
\end{equation}
Both metrics trivially match on the surface $r=r_{0}$ with the sole condition of fixing the mass $M$ of the exterior solution to $M=4 \pi \varrho_{0} \, r_{0}^3/3$. However, the underlying, proper tetrad fields representing the corresponding solutions within the context of $f(T)$ gravity, are quite intricate.

Let us briefly revisit the proper frames leading to the exterior Schwarzschild metric first; full details can be found in \cite{Nossphe}. In what follows, isotropic coordinates with \emph{radial} marker $\rho$ will be used, and the isotropic radius of the star will be $\rho_{\star}$ (not to be confused with $\varrho_{0}$, the constant energy density of the star). The analysis starts by considering the \emph{asymptotic frame}

\begin{equation}\label{asym}
e^t=A_{+}dt,\,\,\,e^{i}=B_{+}dx^i,
\end{equation}
being $A_{+}=A_{+}(\rho)$, $B_{+}=B_{+}(\rho)$. This tetrad leads to the Schwarzschild metric in isotropic coordinates $ds^{2}_{+} = A_{+}^2 dt^{2} - B_{+}^2 (d\rho^2+\rho^2d\Omega^2)$, where $\rho^2=(x^1)^2+(x^2)^2+(x^3)^2$, and

\begin{equation}\label{schis}
 A_{+}=\frac{2\rho-M}{2\rho+M}, \,\,\, B_{+}=\left(1+\frac{M}{2\rho}\right)^2.
\end{equation}
The tetrad (\ref{asym}) is not a consistent solution of the $f(T)$ equations of motion; however, we can radially boost it in order to achieve a null Weitzenb\"{o}ck invariant $T$. The so obtained tetrad automatically will solve the equations of motion for any ultraviolet $f(T)$ deformation, i.e., for any function verifying $f(0)=0$ and $f'(0)=1$ (see eqs. (\ref{ecuaciones})). After the $\rho$-dependent boost  $e_{+}^{\,\,a}=\Lambda^{\,\,\,a}_{+\,b}(\rho)e^b$, the \emph{Schwarzschild frame} $e_{+}^{\,\,a}$ corresponding to the exterior solution results in

\begin{align}\label{schframe}
 e_{+}^{\,\,t}&= A_{+} \gamma_{+} \,dt-B_{+} \Gamma(\gamma_{+})d\rho,\\
 e_{+}^{\,\,1}&=- A_{+}\Gamma(\gamma_{+})\sin\theta\cos\phi\, dt+B_{+}\gamma_{+}\,\sin\theta\cos\phi\, d\rho\,+
 \notag\\
 & +\rho \,B_{+}\,[\cos\theta\cos\phi\, d\theta-\sin\theta\sin\phi \,d\phi],
 \notag\\
  e_{+}^{\,\,2}&=- A_{+} \Gamma(\gamma_{+})\sin\theta\sin\phi\, dt+B_{+}\gamma_{+}\,\sin\theta\sin\phi\, d\rho\,+
  \notag\\
 & +\rho \,B_{+}\,[\cos\theta\sin\phi\, d\theta+\sin\theta\cos\phi \,d\phi],
 \notag\\
  e_{+}^{\,\,3}&=- A_{+}\Gamma(\gamma_{+})\cos\theta\, dt+B_{+}[\gamma_{+}\,\cos\theta d\rho-\rho\sin\theta\, d\theta]. \notag
 \end{align}
 Here we have defined $\Gamma(\gamma_{+})=\sqrt{\gamma_{+}^2-1}$, where $\gamma_{+}=\gamma_{+}(\rho)=1/\sqrt{1-\beta_{+}(\rho)^2}$, and $\beta_{+}(\rho)$ is the non-dimensional boost speed. It was demonstrated in \cite{Nossphe} that the Weitzenb\"{o}ck invariant is null provided the boost is tuned according to
\begin{equation}\label{boostpar}
 \gamma_{+}(\rho)=\frac{M^2+4\rho^2+k_{+} \rho}{4 \rho^2-M^2},
\end{equation}
where $k_{+}\geq0$ is a free constant with units of length \footnote{In ref. \cite{Nossphe} this constant was not considered.}. The condition $k_{+}\geq0$ is necessary in order to have $\gamma_{+}\geq1$, hence $\Gamma(\gamma_{+})\geq0$. Note that the speed $\beta_{+}(\rho)$ goes to one when $\rho\rightarrow M/2$, which corresponds to the event horizon in the isotropic chart $(t,\rho,\theta,\phi)$, which is related to the standard Schwarzschild coordinates $(t,r,\theta,\phi)$ by the radial scaling  $2\rho=r[\sqrt{1-2M/r}+1]-M$. This is actually a deficiency of the isotropic chart, and it has no physical consequences because the junction surface corresponding to the (isotropic) radius of the star $\rho_{\star}$ will always be greater than $M/2$. Moreover, $\beta_{+}(\rho)$ goes to zero when $\rho\rightarrow \infty$, justifying the name adopted for the tetrad (\ref{asym}).

The tetrad (\ref{schframe}) with the condition (\ref{boostpar}) represents the exterior Schwarzschild geometry in any $f(T)$ deformation of ultraviolet character. The free constant $k_{+}$ appearing in (\ref{boostpar}) will be crucial in the matching procedure to be unfolded in a moment. It actually defines a 1-parameter subgroup of Lorentz boosts in the radial direction $\rho$, leaving unchanged the null value of the Weitzenb\"{o}ck invariant $T$; then, it provides valuable information concerning the remnant group of the Schwarzschild exterior solution.

\bigskip
Let us proceed now to fully derive the tetrad field corresponding to the interior metric (\ref{schain}). As before, we start from the isotropic tetrad $e^t=A_{-}dt,\,\,\,e^{i}=B_{-}dx^i$, thus $ds^{2}_{-} = A_{-}^2 dt^{2} - B_{-}^2 (d\rho^2+\rho^2d\Omega^2)$, where now we have

\begin{align}\label{schin}
 A_{-}&=\frac{4 \rho_{\star}^3(\rho_{\star}-M)+M(4\rho_{\star}-M)\rho^2}{(2\rho_{\star}+M)(2\rho_{\star}^3+M\rho^2)},\notag\\
 B_{-}&=\frac{1}{4}\frac{(2\rho_{\star}+M)^3}{2\rho_{\star}^3+M \rho^2}.
\end{align}
It is straightforward to show that (\ref{schis}) and (\ref{schin}) $\mathcal{C}^1$-match at the junction value $\rho=\rho_{\star}$. Due to the fact that we are using isotropic coordinates, the mass $M$ in (\ref{schin}) is related to the constant energy density $\varrho_{0}$, but in a tricky way. In fact we have

\begin{equation}\label{relrad}
y=\frac{4x}{(1+x)^6},\,\,\,\,\, x=\frac{M}{2\rho_{\star}},\,\,\,\,\,y=\frac{\rho_{\star}^2}{R^2}= \frac{8}{3}\pi \rho_{\star}^2\varrho_{0}.
\end{equation}
Note that if we demand $M=0$ (i.e., $x=0$), we have $y=0$, or equivalently $\varrho_{0}=0$ for any $\rho_{\star}$. Under these circumstances we have $A_{-}=B_{-}=1$ from (\ref{schin}) and the metric is just $ds^{2}_{-} = dt^{2} - (d\rho^2+\rho^2d\Omega^2)$, which represents the Minkowski line element in \emph{usual} spherical coordinates; isotropic and spherical coordinates coincide when $M=0$, so we shall take $M\neq0$ from now on. Details concerning the structure of the interior Schwarzschild solution in isotropic coordinates can be found in \cite{Wyman}.

In the hope of finding the interior solution in the $f(T)$ context, we radially boost the isotropic frame $e_{-}^{\,\,a}=\Lambda^{\,\,\,a}_{-\,b}(\rho)e^b$. Of course, we obtain an expression formally identical to (\ref{schframe}), namely
\begin{align}\label{schframin}
 e_{-}^{\,\,t}&= A_{-} \gamma_{-} \,dt-B_{-} \Gamma(\gamma_{-})d\rho,\\
 e_{-}^{\,\,1}&=- A_{-}\Gamma(\gamma_{-})\sin\theta\cos\phi\, dt+B_{-}\gamma_{-}\,\sin\theta\cos\phi\, d\rho\,+
 \notag\\
 & +\rho \,B_{-}\,[\cos\theta\cos\phi\, d\theta-\sin\theta\sin\phi \,d\phi],
 \notag\\
  e_{-}^{\,\,2}&=- A_{-} \Gamma(\gamma_{-})\sin\theta\sin\phi\, dt+B_{-}\gamma_{-}\,\sin\theta\sin\phi\, d\rho\,+
  \notag\\
 & +\rho \,B_{-}\,[\cos\theta\sin\phi\, d\theta+\sin\theta\cos\phi \,d\phi],
 \notag\\
  e_{-}^{\,\,3}&=- A_{-}\Gamma(\gamma_{-})\cos\theta\, dt+B_{-}[\gamma_{-}\,\cos\theta d\rho-\rho\sin\theta\, d\theta]. \notag
 \end{align}
However, this tetrad will be a solution of the $f(T)$ equations only if we can assure that the intervening Weitzenb\"{o}ck invariant $T$ vanishes. A lengthy but otherwise standard calculation gives

\begin{equation}\label{elT}
 T(\rho)=\frac{32[F(\rho)+G(\rho)\,\gamma_{-}+H(\rho)\gamma_{-}']}{\rho^2(M+2\rho_{\star})^6I(\rho)},
\end{equation}
where now the prime refers to differentiation with respect to $\rho$, and the functions $F,G,H,I$ are

\begin{align}\label{funct}
F&= (M\rho^2-2\rho_{\star}^3)\,\times \notag\\
&[M^2\rho^4(M-4\rho_{\star})+10M\rho^2\rho_{\star}^3(M+2\rho_{\star})-8\rho_{\star}^6(M-\rho_{\star})],\notag\\
G&=(M\rho^2+2\rho_{\star}^3)\,\times \notag\\
&[M^2\rho^4(M-4\rho_{\star})+6M\rho^2\rho_{\star}^3(M+2\rho_{\star})+8\rho_{\star}^6(\rho_{\star}-M)],\notag\\
H&=-\rho(M\rho^2+2\rho_{\star}^3)^2[M\rho^2(M-4\rho_{\star})+4\rho_{\star}^3(M-\rho_{\star})], \notag\\
I&=M\rho^2(M-4\rho_{\star})+4\rho_{\star}^3(M-\rho_{\star}).
 \end{align}
The condition $T=0$ fortunately can be worked out because of the fact that $T$ contains just first derivatives of the tetrad. This permits us to integrate (\ref{elT}) for $\gamma_{-}$; nonetheless, we have to introduce further functions in order to write down the result in a sufficiently legible form. It yields

\begin{equation}\label{boost}
 \gamma_{-}(\rho)=\frac{(M\rho^2+2\rho_{\star}^3)^2}{8\rho\, I(\rho)} [-J(\rho)+K(\rho)+8k_{-}],
\end{equation}
where $k_{-}$ is an integration constant (with units of inverse length) and the two extra functions are
\begin{align}\label{functmas}
J&= \frac{2\rho(M\rho^2-2\rho_{\star}^3)}{(M\rho^2+2\rho_{\star}^3)^3}[M\rho^2(M-10\rho_{\star})+2\rho_{\star}^{3}(5M-2\rho_{\star})],\notag\\
K&= \frac{3\sqrt{2}(M-2\rho_{\star})\arctan(\sqrt{M}\rho/\sqrt{2}\,\rho_{\star}^{3/2})}{\sqrt{M}\rho_{\star}^{3/2}}.
\end{align}
The tetrad (\ref{schframin}) with the so obtained $\gamma_{-}$ is the one correctly describing the Schwarzschild interior solution for any ``high energy'' $f(T)$ theory. As in the exterior case, it depends on a free parameter representative of the 1-dimensional remnant group of (restricted) boosts in the $\rho$ coordinate.

Due to the fact that the functions $A_{\pm}$ (on one hand) and $B_{\pm}$ (on the other), $\mathcal{C}^1$-match on the junction surface $\rho=\rho_{\star}$, the $\mathcal{C}^1$-matching of the frames (\ref{schframe}) and (\ref{schframin}) will be guaranteed if the corresponding boosts join smoothly at $\rho=\rho_{\star}$. By imposing $\gamma_{-}(\rho_{\star})=\gamma_{+}(\rho_{\star})$ and  $\gamma_{-}'(\rho_{\star})=\gamma_{+}'(\rho_{\star})$ we obtain just \emph{one} equation linking the free constants $k_{+}$ and $k_{-}$, namely

\begin{equation}\label{juntura}
k_{+}= -\frac{(M+2\rho_{\star})^2}{\rho_{\star}}\left[k_{-}\rho_{\star}+D(x)\right],
\end{equation}
where the function $D(x)$ reads

\begin{equation}\label{juntura2}
D(x)=\frac{1}{4}\left[\frac{3(x-1)}{\sqrt{x}}\arctan(\sqrt{x})+\frac{3x^{2}+2x+3}{(1+x)^2}\right].\notag
\end{equation}
This function is positive in the domain $(0,1)$, where $x=M/2\rho_{\star}$ is defined. $D(x)$ ranges in the interval $(0,1/2)$ and it goes to zero when $x\rightarrow0$ and tends to $1/2$ when $x\rightarrow1$. Bearing in mind that $k_{+}\geq0$, we notice from (\ref{juntura}) that $k_{-}$ is strictly negative.

\begin{figure}[!htb]
\centering
\includegraphics[width=1.1\columnwidth]{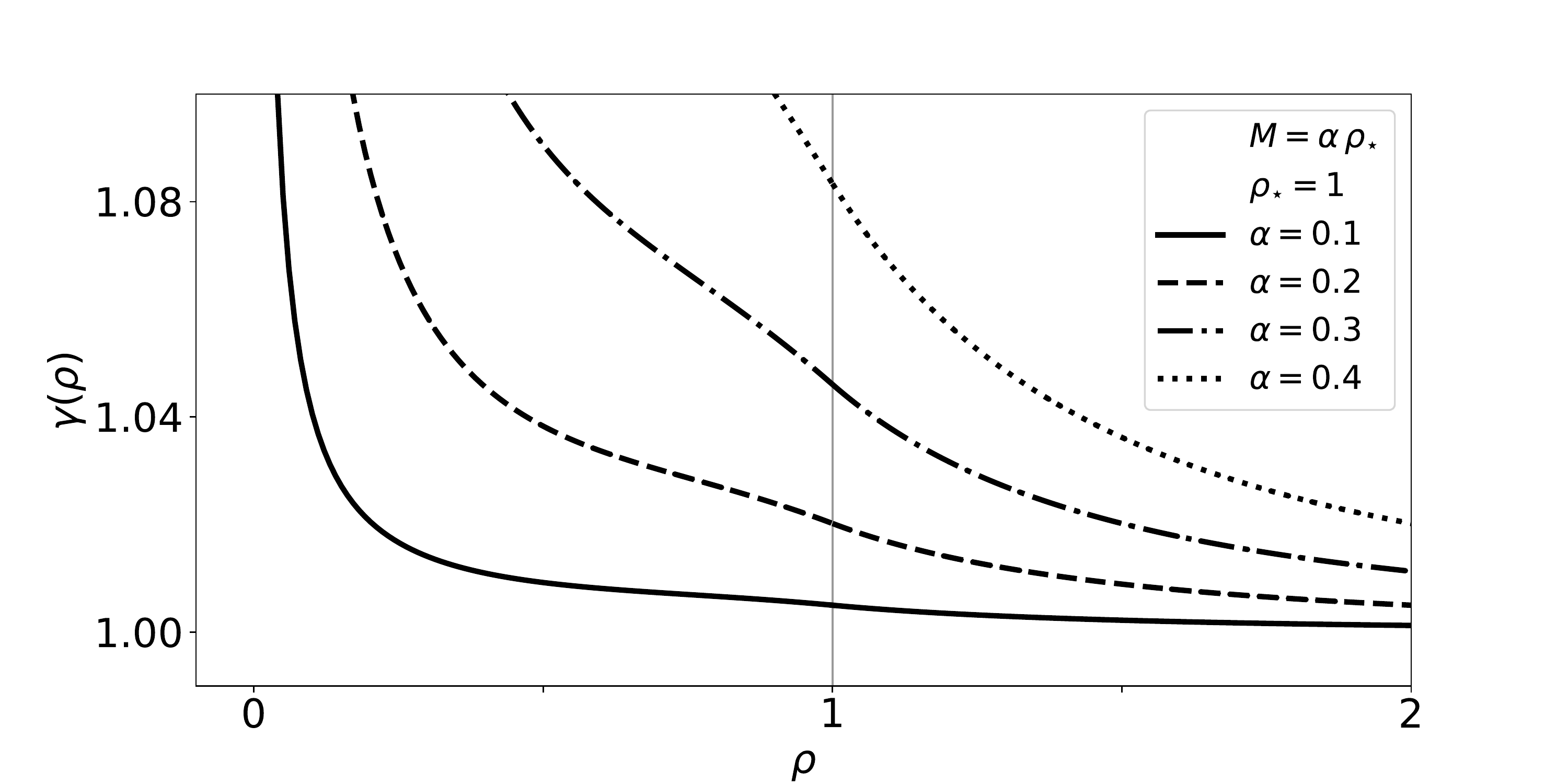}
\caption{$\gamma(\rho)=\gamma(\rho)_{-}+\gamma(\rho)_{+}$, as it comes from Eqs. (\ref{boostpar}), (\ref{boost}) and (\ref{juntura}). Different values of the mass $M$ were taken, for a star radius $\rho_{\star}=1$ (vertical line in the figure).}
\label{fig:gamas}
\end{figure}

Tetrads (\ref{schframe}) and (\ref{schframin}) with $\gamma's$ (\ref{boostpar}) and (\ref{boost}), along with the relation (\ref{juntura}), are the final solution for this particular matching problem. In Figure \ref{fig:gamas} we plot $\gamma(\rho)=\gamma(\rho)_{-}+\gamma(\rho)_{+}$ as it emerges from our analysis, having fixed $\rho_{\star}=1$, for several values of the mass $M$ in agreement with the condition $\rho_{\star}>M/2$.

\section{Closing remarks}\label{remarks}

The procedure underlying the matching of different solutions within the context of $f(T)$ was discussed. Clearly, it does not involve only the matching of the metric components, but of the entire tetrad field instead. This fact brings additional complications at the time of obtaining global solutions in theories possessing absolute parallelism, where the tetrad structure is commonly quite intricate. Throughout this work we emphasized the need for knowing (at least partially), the remnant local symmetries acting on a given solution of the $f(T)$ equations of motion. By examining the remnant local transformations allowed in the intervening tetrads, we exemplified the matching of the Schwarzschild interior and exterior spacetimes representing a simple model of a spherically symmetric star.

Due to the high symmetry of the spaces considered, simple radial (local) boosts acting on the isotropic, diagonal tetrads, were enough to achieve the condition $T=0$ everywhere, assuring that the corresponding spaces are indeed solutions of the $f(T)$ equations of motion for \emph{any} (at least twice differentiable) $f(T)$ function of ultraviolet character. The freedom arising by the presence of an arbitrary constant (let us say, $k_{-}$) in (\ref{juntura}) is probably related to the nature of the additional degrees of freedom present in $f(T)$ gravity, a subtle point not fully understood at the moment \cite{RafaMajo1}. This is analogue to what happened in the example of section (\ref{sec2}) concerning Minkowski spacetime; there we had an infinite number of boosts $\theta(t,\textbf{x})$ able to smoothly glue the tetrads across the surfaces $t=\pm1$.

Nonetheless, as mentioned at the end of the preceding section, we found that $k_{-}$ must be negative in order for $k_{+}$ to be positive. This implies, by Eq. (\ref{boost}), that $\gamma_{-}$ is divergent at $\rho=0$ due to the $\rho^{-1}$ factor present there (note that $I(\rho)$ is non null at the origin, and both, $J(\rho)$ and $K(\rho)$ tend to zero as $\rho\rightarrow0$, see Eqs. (\ref{funct}) and (\ref{functmas}) respectively). Hence, regardless of the regularity of the functions $A_{-}$ and $B_{-}$ at the origin (and of the Weitzenb\"{o}ck scalar, which vanishes everywhere), the tetrad field becomes singular there. Ultimately, this is a consequence of the fact that the boost speed becomes the speed of light at the origin. On the other hand this behavior is somewhat expected: the tetrad is constituted by four 1-form (or vector) fields, hence it should be null or divergent at the origin of the spherical coordinate system held at the center of the star, the former being excluded by the constant (non null) character of some metric components at the origin.

It is not hard to foresee that, under more general circumstances, the matching procedure will become rapidly more complex and less intuitive. For instance, if we would deal with the collapse of a dust cloud and the subsequent black hole formation within the realm of $f(T)$ gravity --i.e., the analogue of the Oppenheimer-Snyder solution in GR-- it is clear that the local symmetries to be considered should be time-dependent. Additionally, due to the fact that the spaces to be matched in this case have a different number of isometries, it results plausible that we would need to consider, in addition to boosts, also time-dependent rotations. All the more reason, a similar complexity will arise when deformed solutions be considered (i.e. solutions of $f(T)$ gravity which are not present in GR), as it is the case, for instance, of regular cosmological models \cite{FF} and black holes \cite{CQG}. The comprehension of the dynamics of the tetrad field in situations of this sort is certainly a good motivation for continuing with this line of research. In particular, it would be pertinent to investigate the appearance of further constraints coming from the matching procedure which might have an impact on the asymptotic structure of the tetrad field, specially in the case of asymptotically flat spacetimes, where the nature of the additional degree/s of freedom seems to be slightly more unmaskable \cite{Golovnev},\cite{Beltran}.

\vspace{-0.5cm}
\subsection*{Acknowledgments}
FF is a member of Carrera del Investigador Científico (CONICET), and his work is supported by CONICET and Instituto Balseiro. MO is a PhD student supported by CONICET.

\end{document}